# Privacy protection based on mask template

Hao Wang,[1] Yu Bai,[2] Guangmin Sun,[3] and Jie Liu[3]

[1] Network and Information Technology Centre, Beijing University of Technology, Beijing 100124, China.
[2] Medical Engineering Division, Beijing Friendship Hospital, Beijing 100050, China.
[3] Information Department, Beijing University of Technology, Beijing 100124, China.

Correspondence should be addressed to Hao Wang; wanghao@bjut.edu.cn

## Abstract

Powerful recognition algorithms are widely used in the Internet or important medical systems, which poses a serious threat to personal privacy. Although the law provides for diversity protection, e.g. The General Data Protection Regulation (GDPR) in Europe and Articles 1032 to 1039 of the civil code in China. However, as an important privacy disclosure event, biometric data is often hidden, which is difficult for the owner to detect and trace to the source. Human biometrics generally exist in images. In order to avoid the disclosure of personal privacy, we should prevent unauthorized recognition algorithms from acquiring the real features of the original image.

In this paper, we take face data as an example and propose the mask template application of protecting the privacy in images, which can help the original image establish stealth effect. This technology can blind the target recognition algorithm at the pixel level by generating distributed mask model and imperceptible perturbation respectively. Noise pixels are added to the original image through the mask template to confuse the target recognition algorithm, and the image that has lost the feature focus can still be recognized by the human eyes, so as to protect the privacy in the image, and authorized users use mask templates to restore protected images. Using this method, multiple identification algorithms of the same type can be superimposed to expand the compatibility of protection. The mask template can be used as a key to restore the image completely which also can be encrypted by conventional methods to further prevent cracking.

## Introduction

Feature recognition applications based on artificial intelligence, especially the proliferation of face recognition models, pose an unprecedented threat to personal privacy. It is common for recognizers to obtain personal face data without consent. As early as at the "China 315 party" in 2017, the host logged in to a correspondent's account by forging his face. On February 27, 2020 US time, CLEARVIEW AI was attacked by hackers[1], and over 3 billion face data and identity information were leaked, including the data of 600 law enforcement agencies such as the US immigration service, the Department of justice and the Federal Bureau of Investigation. The owners of these data can view anyone's basic information, social relations and criminal records, causing immeasurable losses to individuals and society. Twining launched by POPSUGAR is favoured by users all over the world. Users only need to upload their own photos to find stars who look similar to themselves. Unfortunately, it also has a serious disclosure of user privacy. The event of "Buy 70 stars ID photos only need RMB 2" was also caused by the disclosure of personal privacy in the JIAN KANG BAO applet.

The development process of biometric recognition technology is in full swing. From 2010 to 2018, the average annual compound growth rate of face recognition market alone is 30.7%, and the market scale will exceed 10 billion yuan by 2024. Such a development speed makes our faces nowhere to hide. The accuracy of face recognition algorithms represented by MEGVII, ArcSoft, SenseTime, DahuaTech and other manufacturers exceeds 99.99%, which are widely used in community access control, shopping malls, monitoring and public security systems. By August 2019, the number of surveillance cameras in China has reached 176 million, and 450 million cameras will be newly installed by 2020. The "smart city" and "bright project" have further stimulated the demand for face recognition algorithms and high-definition cameras. 5G technology has also increased the transmission capacity, resulting in disruptive reform in the whole industry. Feature technology recognition is not only accurate but also rapid. The convenience of biometric authentication technology makes people's memory gradually "degenerate into a goldfish". When iPhone x released face unlocking, it was once resisted by netizens, worried about the safety of face data, and even claimed to bring a "Face Bikini" when going out. Now many users have forgotten the "face leakage" and enjoy the convenience of face unlocking like frogs in the warm water. From a technical point of view, it is difficult for people to protect their privacy through their own consciousness.

Unfortunately, the current research on privacy protection in image recognition is lack of effectiveness. Some methods of distorting or confusing images have been widely studied. Although Rubik's cube and dynamic password (RCDP) [2] algorithm is difficult to crack, human beings cannot understand the encrypted image. For the image owner, the confidentiality value is higher than the application value, and cannot be used in front-line medical treatment and researchers. Fawkes[3] adds the disturbance to the key points of the face image through the cloak, which can effectively interfere with some recognition algorithm, but this method can neither restore the image nor prevent the cracking of generative adversarial networks(GAN). Therefore, a system or method that can encrypt the original image is needed to prevent unauthorized recognition algorithm from reading the image, and the protected image does not affect human and authorized recognition algorithm. It maintains the status quo of existing identification and storage equipment.

This paper introduces the relevant works of avoiding privacy leakage, then discusses how Mask Template Network and Perturbation Generation Network protect privacy in images, finally verifies the preciseness and availability of the method through experiments. The innovation of this paper is that a general attack model can be trained to encrypt the image, and the image can be restored to the state before encryption through mask.

## Review of Literature

In privacy protection, backdoor attacks[4] are usually used to cheat the recognizer. The attack methods for data and feature are Trojan Horse Attack[5]，Data Poisoning Attack[6], Clean Label Attack[7], Pruning-Aware Attack[8]. Methods for IO and test sets are One pixel poisoning attack[9]， StingRay Attacks[10], PATOM Multi-Task attack[11], Data Poisoning Against Differentially-Private Learners[12] , etc. The attack methods for the model are Zero-Bit Watermarking Algorithm[13], Regular Embedding Watermarks Algorithm[14], DeepMarks[15], BadNets[16], PoTrojan[17], Poisoning Attacks to Graph-Based Recommender Systems[18], Indirect Adversarial Attacks via Poisoning Neighbors for Graph Convolutional Networks[19] , etc. In the changing era of artificial intelligence, there have been defensive schemes. The major countermeasures are shown in Table 1.

Table 1: Countermeasures of privacy protection.

| Sort | Case | Theory | Modify model |
|---|---|---|---|
| Data and feature modification | Data pre-processing[20] | Pre-process input data | NO |
| | AUROR[21] | Automatically identify and display the shielding characteristics of abnormal distribution | NO |
| Model modification | Pruning[22] | Eliminate the dormant neurons on the pure input to reduce the size of the backdoor network | YES |
| | Fine-pruning[22] | Trim dormant neurons and fine-tune the model | YES |
| | Fine-tuning[24] | Train the poisoned neural network to make the backdoor trigger invalid | YES |
| | DeepInspect[25] | The conditional generation model is used to learn the probability distribution of potential triggers from the query model | YES |
| Output defence | Loss based defence[26] | If the target model loss exceeds the threshold multiple times, an accuracy check will be triggered | NO |
| | Integrated defence[27] | Combined with the prediction results of different models to judge the prediction category of samples | NO |
| | Detector defence[28] | The input is detected by support vector machine and decision tree | NO |
| | Multi task model defence[29] | Improve robustness through data cleaning and multitasking learning | NO |

When these traditional attacks are applied in privacy protection of images, they have a visual impact on the original image, and may even cause serious distortion of the image. If these poisoned images are used directly, they will completely lose their visual reference value and cannot be displayed in some applications, such as certificate photos, patient electronic documents, etc.

The algorithm proposed in this paper aims to disturb the classification mechanism of the recognizer, and can restore image through the key without affecting the visual effect. After decryption, the protected image has the same feature distribution as the original image, which can be re-applied to the authorized recognition algorithm.

The implementation features include: (a) The encryption network does not mis-classify a single pre-selected image, but classifies the current and future inputs with the same features at the model level. (b) In order to ensure the immunity of different recognition algorithms, the attack model does not need to be trained from the beginning. (c) Prevent anomaly detection in feature space and make the attack network output diverse and random. (d) The protected image can be restored to the state before poisoning in the process of propagation.

## Methods

This chapter proposes two networks to protect the target image and explains the restoration method of the protected image.

The traditional method of deleting labels is applied in some special fields, which is not conducive to tracking the research object, because the common poisoning attack will block the key features of the image. The anomaly detection of some recognition algorithms will return the poisoning identification, so that the person using the data cannot identify the target in the image. To prevent the human eyes and detection algorithm from noticing that the image has been modified, we need to add imperceptible interference to the key pixels of the image. In practical application, this can not only prevent unauthorized recognizers from collecting data but also let researchers or users clarify the subject. From a mathematical point of view, the disturbed image is two distributions. Sometimes we need to restore the disturbed image in order to apply it to officially authorized batch recognition. For example, toddler photos in hospitals are protected, but for the needs of scientific research, different medical institutions need to share their data to predict the relationship between blood type and eye colour. As long as the hospital's ethics committee passes the review and is approved by the photo owner, the data may be shared and used. However, the risk of privacy disclosure is often ignored and difficult to control by both parties. It is necessary to desensitize the original photos to ensure that the circulating data is not obtained by unauthorized recognition algorithms, and to ensure that researchers can observe and apply images normally. After the data classification, face recognition may map the identity information of infants. It is obvious that the tampered image cannot be successfully matched with a real person. This requires us to restore the changed content, remove the interference factors by using password or mask, get the original image, and then carry out subsequent matters. We will use Mask Template Network and Perturbation Generation Network to complete this challenge.

**Mask Template Network**

Major function of this network is to train a mask template model can generate the corresponding feature recognition algorithm. The output of the model contains random factors about the distribution of feature points corresponding to the target algorithm. The accuracy of the feature recognition algorithm can be affected by coupling the random factors to the corresponding pixels of the original image. Original image can be restored by printing these random factors and calculating with protected image. According to this idea, we assume that the regression function $f$ is the feature point mapping learned in a recognition algorithm. In the training stage, the same mask template distribution needs to be got from images of different scales, and the complete feature space should comply with Equation 1. Vector $\boldsymbol{\varphi}$ is the $n_{\text{th}}$ feature distribution of different scales ($s$) in the original images. To make the network to learn the hyper parameters, Equation 2 only needs to be established. The relatively simple idea is translation and scaling. Suppose that the aim function is Equation 3.

$$\boldsymbol{\phi} = \sum_{1}^{n} \boldsymbol{\varphi}_n^s \qquad (1)$$

$$f(\boldsymbol{\varphi}_n^s) = f(\boldsymbol{\varphi}_n^{s\prime}) \qquad (2)$$

$$f_*(\boldsymbol{\phi}) = \omega_*^T \cdot \boldsymbol{I} \qquad (3)$$

$\boldsymbol{I}$ represents input eigenvector, $\omega_*$ is hyper parameter, $*$ refers regional coordinate of the feature point, $f_*$ is predicted value, which should have the smallest Euclidean distance from the real value $t_*$, the loss function can be defined as Equation 4.

$$Loss = \sum_{i}^{m} (t_*^i - \omega_*^T \cdot \boldsymbol{I}^i)^2 \qquad (4)$$

The function optimization objective is defined as Equation 5.

$$W_* = argmin_{\omega_*} \sum_i^m (t_*^i - \omega_*^T \cdot I^i)^2 + \lambda \|\omega_*\|^2 \tag{5}$$

Pictures with landmarks are classified according to the face recognition algorithm version on the starting points of the training process. Images at input end are samples of different scales and distinct faces which are mapped to a fixed size landmark template image, and the feature output points are given by the same recognition algorithm. The feature points region output by the algorithm is pre-processed to 4×4×3 pixels or 4×4×1 pixels to reduce the amount of computation (the result of grey scale image is sometimes not ideal, because some algorithms are sensitive to colour channels). Normalizing a random facial mask model is expected to have 30 million super parameters, so it is necessary to establish a simple network. Referring to VGG-16[30], each layer adopts batch normalization layer to speed up model training and prevent over fitting. We use filter of 4 ×4 pixels and same convolution (stride is 2). After 2 consecutive convolutions with 16 filters, the image is compressed to 128×128×32 pixels through the pooling layer. Then there are several convolution layers again, using 64 filters and some same convolution. After pooling, it becomes a 64×64×64 matrix. Subsequently, the matrix data of 16×16×128 pixels are gained through 128 filters and maximum pooling twice respectively. After passing through two full connection layers, it is input into the Softmax layer for activation, as shown in Figure 1. The predicted classification corresponds to the feature point coordinates output by the same human face in the same recognition algorithm.

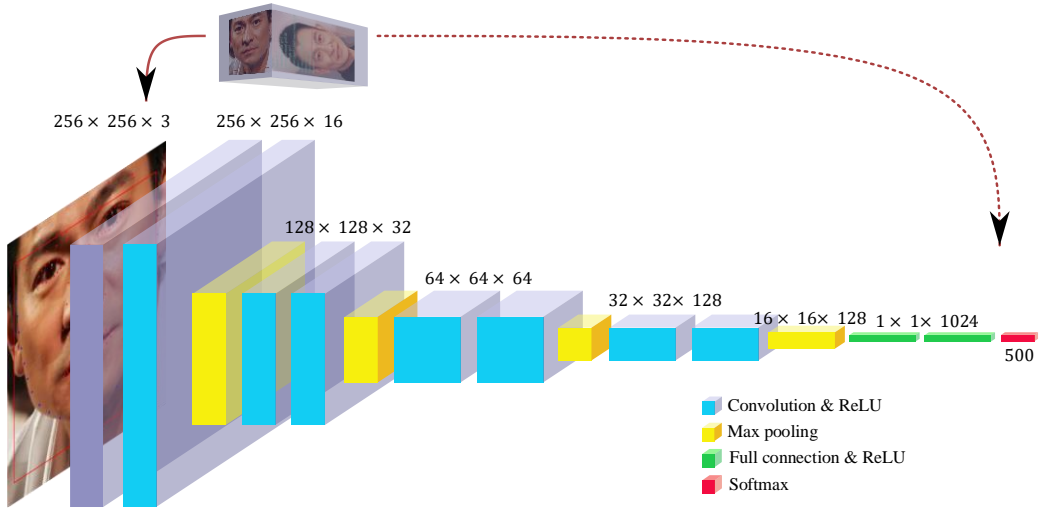

Figure 1 Mask template network

We inject 75% of the data in CelebA dataset into the face recognizer to get the feature points of each image and then do the maximum likelihood operation $f_*(\boldsymbol{\phi}; \theta)$ on the distribution of the feature points . In the formula, the distribution parameter $\theta$'s potential role is to make the Euclidean distance between desired distribution $t_*$ and measurement $f_*$ is as small as possible. If the trained model is universal, Feature-representation-transfer[31] can be used as a node to fit other recognition algorithms, and gradually adapt to more and more face recognition algorithms through continuous iteration and sharing. Machine learning (ML) can extract memory from models[32]. This method can share the details of training dataset and has good reference value, but the problem is that the transfer process is invalid for untrained models. At first, we wanted to transfer the features back layer by layer to deal with the constantly upgraded third-party recognition algorithm, but it produced a negative transfer effect. The solution is to train nodes with different versions of recognition algorithms, save the coordinate features of the obtained feature template and transfer them back. The template features of each round of training are marked with different indexes. When generating the template, attackers can select all the templates or a specific target recognition algorithm according to the actual needs. When the face mask model trained with CelebA dataset is applied to the recognition algorithms of different retail versions, it is reasonable that the tasks should have correlation. However, this commercial retail algorithm belongs to black box application, and there is no way to define the correlation, nor can describe the correlation of each task by mathematical method. Therefore, we need to cluster and classify the vulnerability points of each model before training the mask template to ensure the universality of the attack system. Some recognition algorithms need feature point clipping to output the model correctly.

We came up with an ideal solution, as the feature points of different recognition algorithms are migrated together. For example, the corner of the eye, the corner of the mouth and the tip of the nose as general features should be the main positioning points in each recognition algorithm. If these feature points are aggregated separately, the generated model will inherit the characteristics of all algorithms. A shared low dimensional space is used to reduce the difference between the target domain $x_t$ and the source domain $x_s$ of the models trained for different recognition algorithms, and finally producing a smaller distribution of local features through continuous iteration. The new model generated by the local features of these cross domains can be printed by visual neural network, and the formed image is the basis for reasoning mask template, as shown in Figure 2.

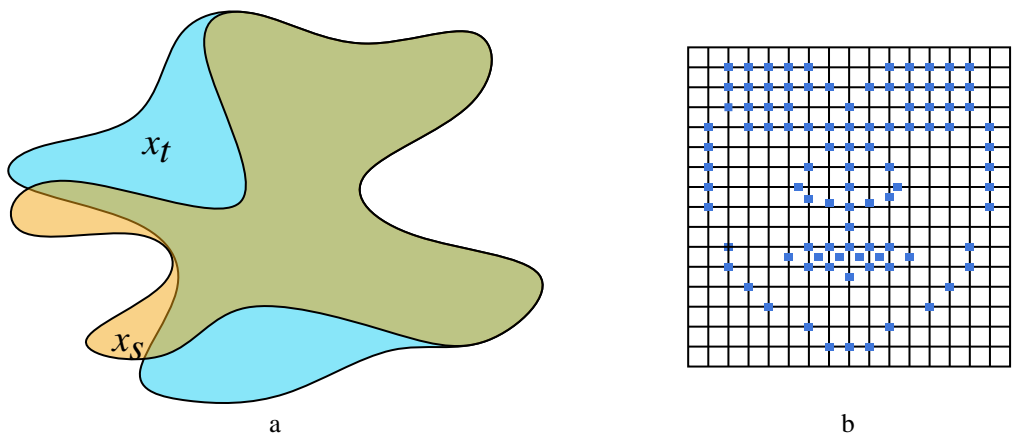

a                                b

Figure 2 Template of overlapping features. (a) Overlapping features of source and target domain. (b) Facial mask template of 2 facial recognition algorithms

In the actual production environments, we can't cluster all the feature points of each recognition algorithm separately, because there will always be some unknown data, and the feature distribution of some algorithms is very different, so the trained model will be very large. We need elements lower than algorithm feature points to represent the region concerned by the current algorithm. If the features of the same category can be mapped to the same low dimensional space, the category features will have similar examples in the high dimensional space. For example, regarding distribution of the nose, we can map the distribution of tip and wing of the nose to one vector. As an adaptation factor, different algorithms have different coefficient matrices **τ** for the vector. The training network can be adapted by manually modifying the coefficients without retraining for the new algorithm. To automatically adapt to the recognition algorithm, only the matrix **τ** linear classification and label assignment are required, as shown in Figure 2.

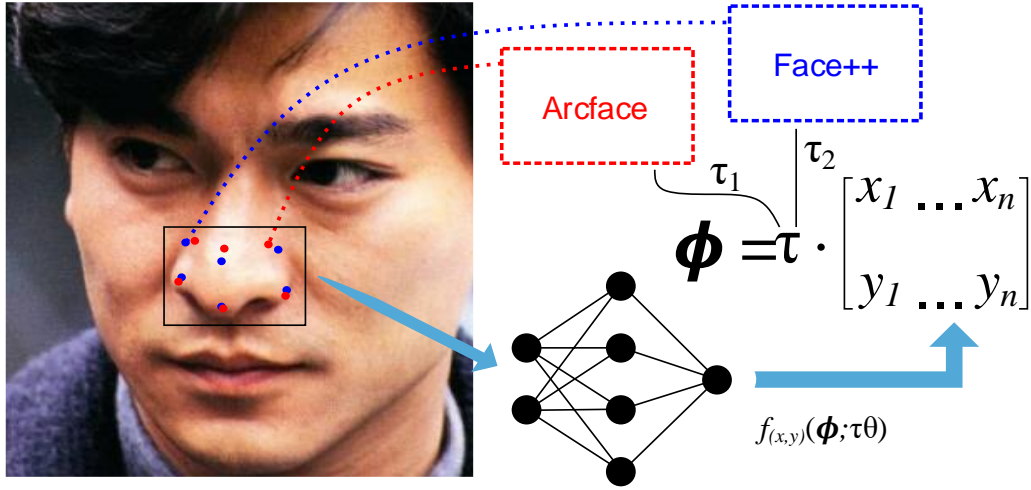

Figure 3 Mapping of τ in different algorithms

If the generated template is unique, just adjust the offset of * in $f_*$ (which can be understood as salt and pepper noise in the encryption function), * is a coordinate set, and the capacity depends on the target recognition algorithm, up to 500. Outputting a proprietary mask template for different users can effectively prevent anomaly detection in the feature space and cracking by other networks.

**Perturbation Generation Network**

In the current research, the principle of our deception recognition algorithm is to change the image feature distribution by adding some cloaks to the image, so that the recognizer or algorithm cannot capture the real feature relationship. These interferences will not affect human eyes browsing and application of images. According to the above requirements, we use numerous of methods and ideas recommended by predecessors for improvement. In the first step, the mask template region of the original image is fitted to a non-linear activation function, which differs completely from the original image. In the second step, if the generated image has undergone significant changes in vision, it is necessary to use noise to generate an image that is confused with the real. The noise should be visually close to the original image. The third step is to fine tune the generation network to verify the recognition effect of the algorithm and human eyes.

We use Gaussian distribution to train the new model and initialize the face image into the mask template of 128×128×3 pixels. Input image $p$ into algorithm $R_l$ should conform to Equation 6, and $l$ is the algorithm version. A feature in the template $\boldsymbol{\varphi_n}$ and disturbance characteristics $\boldsymbol{\widehat{\varphi}_n}$ satisfies perturbation mapping, as in Equation 7.

$$R_l(p) \in f(\boldsymbol{\phi}) \qquad (6)$$

$$\begin{cases} \gamma = \operatorname{argmax}_\gamma Dist(R_l(\boldsymbol{\varphi_n}), R_l(\boldsymbol{\varphi_n} \oplus |\gamma(\boldsymbol{\varphi_n} + \boldsymbol{\widehat{\varphi}_n})|)\tau) \\ \gamma(\boldsymbol{\varphi_n} + \boldsymbol{\widehat{\varphi}_n}) < \sigma \end{cases} \qquad (7)$$

*Dist*(.) means the Euclidean distance of the two characteristic distributions, τ is recognition ability coefficient of the recognition algorithm. σ shows perceptual perturbation budget, it is the weighted average of the current gradient and purports human eyes recognition ability coefficient. γ represents disturbance model. The generated network can apply to the original recognizer. Theoretically, any recognition algorithm can be used as the discriminator in the *f* distribution of the mask template. Of course, we can constantly add and change other face features in the mask template, or use a gradient different from the region as the generator.

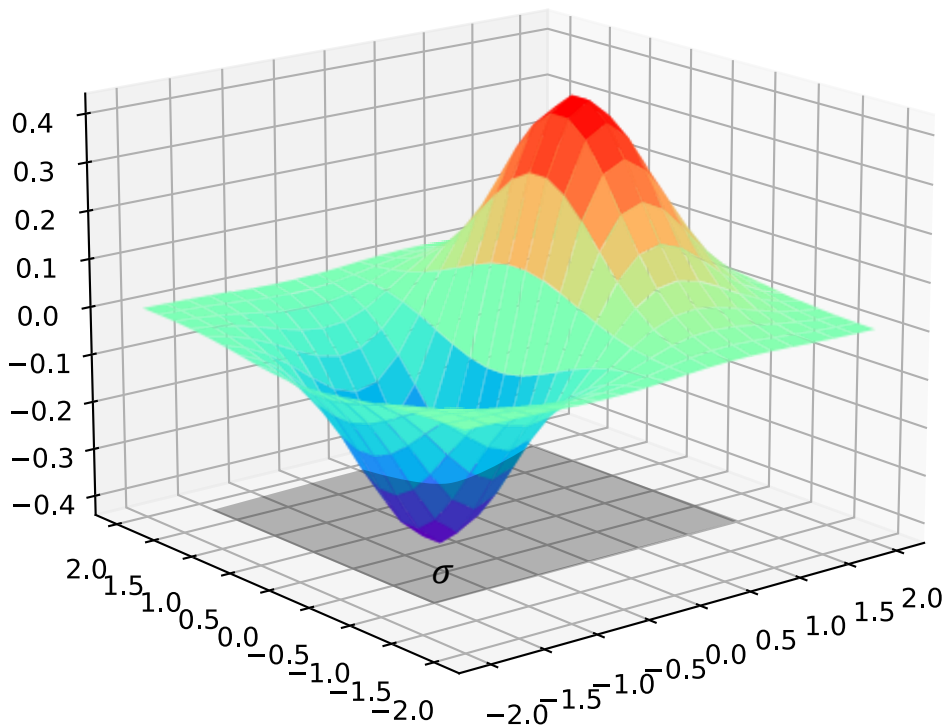

Figure 4 Normalized σ for adjusting human visual threshold

Through continuous training, the recognition rate of discriminator is gradually reduced. So we should constantly fine-tune through observation σ (interception point) plane, control extreme value of γ so that the output image cannot be recognized by the face recognition algorithm, but the human eyes can recognize it until it is satisfied, see Figure 4.

In order for the generator to improve the quality of samples and the speed of convergence, all pooling layers and full connection layers are cancelled. We updated some parameters of DCGAN[33] and added mask template in layer 4, which is mainly to shield non-key feature points and keep the generated image in the original state as much as possible. On the one hand, we can prepare for the restoration of face gradient later, on the other hand, we can reduce the difficulty of adjusting parameter σ and improve the training speed of the network. The difficulty of adjusting parameter σ is that it cannot be adjusted separately for a certain feature point, but to find a balance in all feature point planes, and crossing the boundary will produce a large area of noise pollution.

The fifth layer performs XOR operation with the original face image of fixed size, generates new feature distribution in the corresponding area of the image, and completes the poisoning attack of the output image on the recognition algorithm. See Figure 5 for network view.

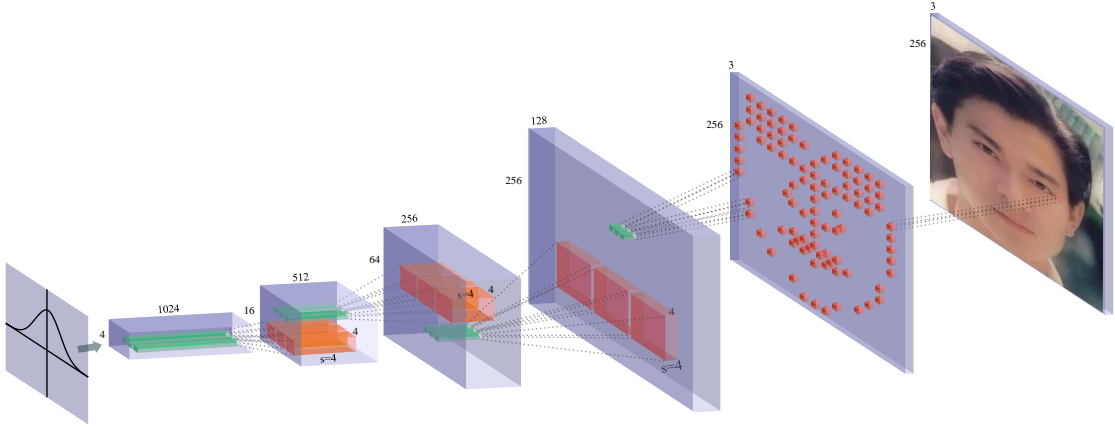

Figure 5 Perturbation generation network based on DCGAN

In generating pictures, we adopt the suggestions of DCGAN author, learning rate at the suggested value of 0.0002, momentum term at the suggested value of 0.5

**Restore image**

After the mask template network reasoning, each user can generate a unique mask template. The protected image is a set of pixels at specific regions generated by random noise filtered by mask template which forms the final appearance only when they are XOR with the face image. As long as we output the mask template generated by this intermediate process into a binary file, it can be used as a key if XOR with the disturbed image again and the pure image can be restored. The contents of mask template can be saved again through image encryption technology [34] to increase its security.

**Experiments**

In this section, we will compare the application effects of several famous privacy protection methods in the current innovative recognition algorithms, and test whether our methods can successfully protect privacy data and restore tampered images. The entire process adopts the black box method, and the evaluation index uses the protection success rate, which is the ratio of the error rate and the accuracy rate of the recognition algorithm for the processed images. $F_c$ represents the probability of the image is incorrectly classified by the recognition algorithm. $T_o$ means the probability of the original image is correctly classified by the recognition algorithm. $T_r$ stands for the confidence of the input recognition algorithm after restoring the picture. They should be referred to Equation 8.

$$r_s = \frac{F_c}{1 + T_o - F_c} \qquad (8)$$

We take the photos of Andy Lau, a famous artist in Hong Kong, as an additional training and test sample because he has quite significant facial features. See Figure 6. Known commercial recognizers or algorithms have a low probability of false recognition. In previous experiments, they have always been our favourite candidate, and his enterprising spirit is also worth learning.

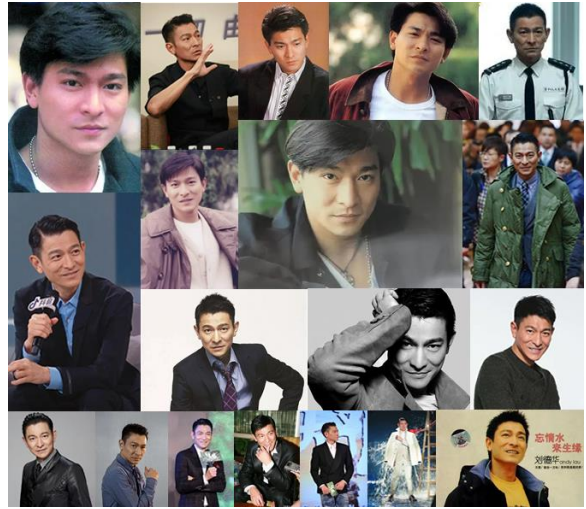

Figure 6 Photos of Andy Lau

For the recognition algorithm tested, we focus on the latest version of faces on Baremetal Compute Service (BCE) of Baidu, which has excellent performance and advanced interference immunity, tamper proof and feature verification functions. BCE algorithm can detect bounding box, face rotation, blur, eye status, emotion, face type, face shape, age, glasses status, gender, level of appearance. Comparison algorithms include Azure face, Face++, ArcFace and open source face recognition algorithm based on Dlib (Dface for short). Figure 7 shows the output of the BEC facial reader written in PHP.

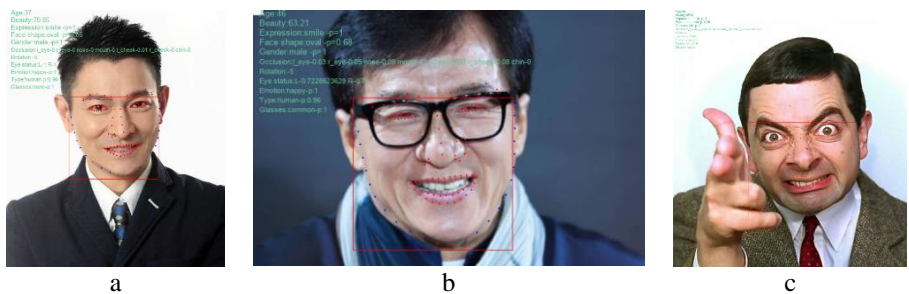

Figure 7 BEC face detector. (a) Andy Lau. (b) Jackie Chan. (c) Rowan Atkinson.

The biggest difficulty in attacking BCE recognition algorithm is that it has a tamper detection network. When the input picture is tampered, it can predict the tampered area according to the relevant feature points, and adjust the threshold of the tampered area to improve the evaluation accuracy. 3D recognition and low resolution recognition technology are also the industry-leading level, and the attack is relatively difficult. Almost all previous privacy protection algorithms have failed in BCE. Even Andy Lau, who blocked his eyes and mouth, was recognized in 200ms, and the average recognition rate was 0.90595. See Figure 8.

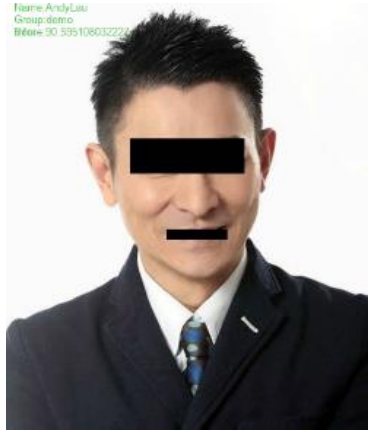

Figure 8 Photo covered eyes and mouth of Andy Lau

**Regional pixel confusion experiment**

The Face Detection and Segmentation(FDS)[35] with Region of Interest Align (RoIA) functions can faithfully preserve exact spatial locations and output the feature map to a fixed size, take the mouth and nose as examples, confuse the pixels in this area, and try to make humans recognize the characters in the picture.

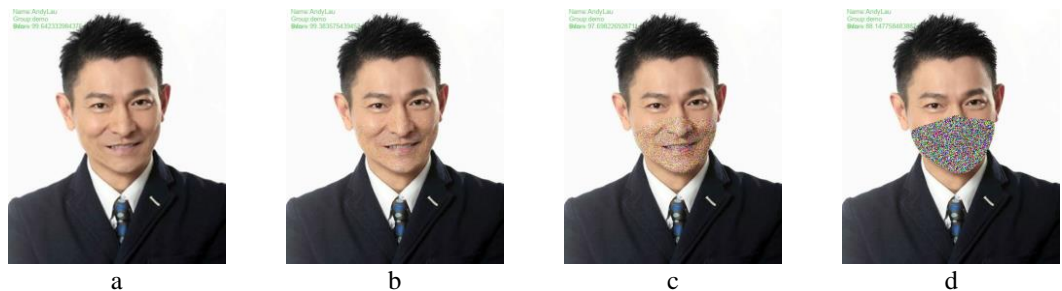

      a            b            c            d

Figure 9 Regional pixel confusion experiment. (a) p=20 $F_c$=0.00358. (b) p=88 $F_c$=0.00616. (c) p=167 $F_c$=0.02302. (d) p=325 $F_c$=0.11852

During the experiment, p represents the confusion threshold, and the greater value, the deeper confusion. The experimental results show that even if the segmented area covers 50%, the recognition rate of BCE is still close to 90%. If it continues to cover near the eyes, the recognition difficulty of Lau by human eyes will be significantly enhanced, and many important facial features will disappear. This method fails in BEC recognition algorithm. See Figure 9.

Using the same method to experiment in Azure face, Face++, ArcFace and Dface, only old version of Face++, ArcFace2.0 and Dface algorithms cannot recognize images with p greater than 167. It shows that this method is outdated

**Feature area twist**

The traditional twist principle is to change the relative position of the target area. Feature points account for only a few to dozens of pixels. It is very limited in the face area, so the local features are extracted first, then the pixel distribution in the limited space is changed, the image feature distribution will be changed. At present, the commonly used feature extraction methods include one-dimensional image feature extraction along the vertical direction of the contour [36] or two-dimensional image feature extraction of the square neighbourhood of feature points[37]. According to twist principle, distorting the distribution of pixels on some feature images can change the image distribution to deceive the recognition algorithm. In the experiment, we simulate the application of directional force on image pixels to force the Euclidean distance of image feature points to change. However, BCE algorithm has the function of tamper area repair, and can predict the original shape of feature space. The degree of the attack image distortion outside the verification ability of tamper proof algorithm is unacceptable to human eyes.

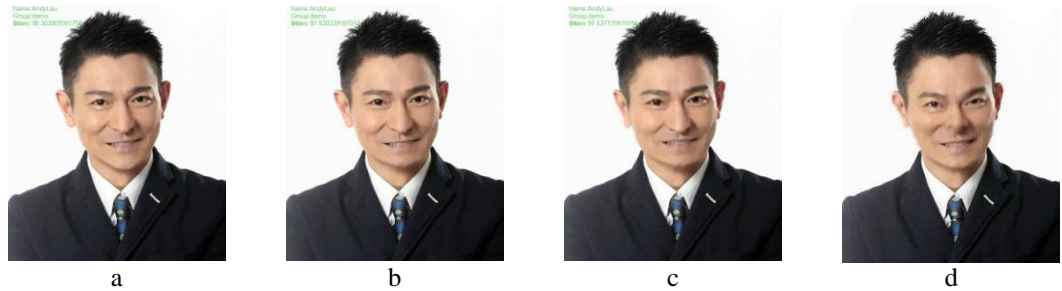

Figure 10 Imperceptible twist attacks samples. (a) Pressure=50 $F_c$=0.01696. (b) Pressure=59 $F_c$=0.0207. (c) Pressure=63 $F_c$=0.02463. (d) Pressure=72 $F_c$=0.804

When the pressure value is 72, the misclassification probability of BEC exceeds 0.8, showing that the anti-recognition attack on the picture has worked, but the handsome Andy Lau is so ugly that can be hardly recognized. See Figure 10. It shows that this method has a certain effect on privacy protection. However, the effect of human eyes recognition on image subject is poor. Besides, Dface(any twist will invalidate it), other algorithms have similar effects. It shows that this method cannot meet the requirements of human eyes recognition and cannot restore the tampered image.

**Fawkes experiments**

The model trained by Shawn Shan[3] can effectively attack ArcFace2.0 and Face++ recognition algorithms. The integrity and fidelity of the portrait are close to the original image, and the human eyes can hardly detect the changed texture on the image. See Figure 11. However, when facing ArcFace 3.0 and BEC algorithm, we took advantage of the training speed advantage of inception-resnet-v4[38] and increased the number of training samples, but the results were seriously distorted or could not reduce the accuracy of target recognition algorithm. To solve the problem of cloak transferability, we use robust feature extractors[39] to reproduce the original experiments. After each algorithm has been tested 500 times by different Structural Dis-similarity Index(SDSI) perturbation budgets, we found that, when the SDSI perturbation budget is less than 0.018, the BCE recognition algorithm is immune, and the recognition accuracy remains above 0.9. When the budget greater than 0.02, the protection success rate exceeds 0.9. See Figure 12 for the results.

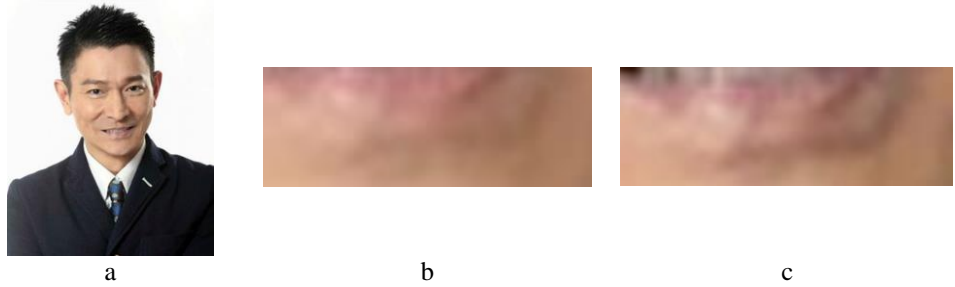

Figure 11 Fawkes attack sample. (a) Images trained by Fawkes. (b) Original picture of Andy Lau's lower lip. (c) Fawke picture of Andy Lau's lower lip

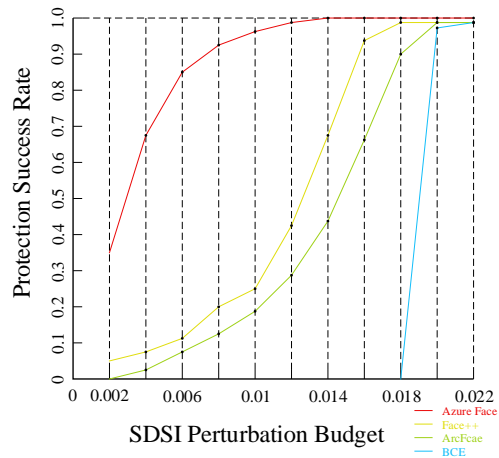

Figure 12 Relationship between parameters and protection rates in Fawkes algorithm

From the output data, it seems to meet the needs of privacy protection, but the printed image content is seriously distorted, and the noise is full of the main feature areas. It is difficult for human eyes to identify the key components of the face in image easily. It can be judged from the image edge that this method can make a new distribution, as shown in Figure 13. This also shows that the sudden change of the blue curve can indeed reduce the accuracy in the recognition algorithm.

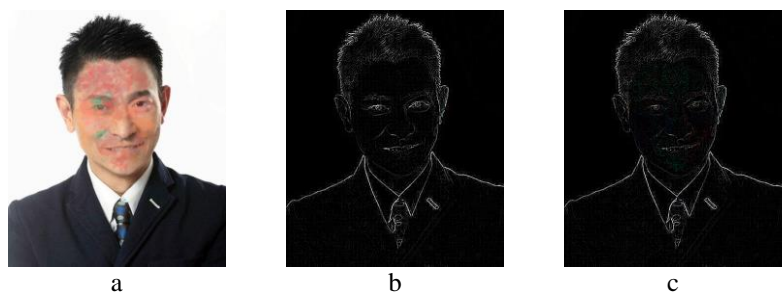

Figure 13 Output image of attacking BCE with Fawkes algorithm. (a) Printout of Fawkes. (b) Original image edge. (c) Fawke image edge

From the analysis results, it can be concluded that Fawkes method can effectively deceive the recognition algorithm, but the processed image has lost value of direct application and cannot meet the needs of multi-algorithm superposition.

**ADVHAT experiments**

The principle of ADVHAT is to train a printable texture[40] that contains features similar to human face, but the distribution of these features differs from the original image. The texture image containing interference was printed on a hat or appearing at the same time with the human face can protect the image by defocusing feature areas location of the recognition algorithm, see Figure 14. We made an exclusive sticker for Andy Lau, which can easily protect the privacy of the characters in the image in ArcFace 2.0 recognition algorithm[41]. The sticker is placed near the forehead, and the characters cannot be accurately identified. Some old algorithms cannot even judge that there is a unique face in the image. Although ADVHAT is very effective for the previous generation of recognition algorithms, it was eliminated only a year later. Now it can hardly attack any recognition algorithms. When the same sample is input into BCE, all features are captured instantly. The average correct recognition rate is 0.9893. Even if the sticker is placed on Lau's eyes, the recognition rate can still be maintained above 0.98. This result is close to the effect of regional pixel confusion experiment. The reason is that the new version of recognition algorithm pays more and more attention to the correlation between feature points, and the collection of feature points is more and more intensive. If it is only confused with similar feature textures, the algorithm can quickly judge that it is not a part of the original image and filter this part of the vectors. Experiments show that ADVHAT cannot attack the new version of the algorithm with feature verification.

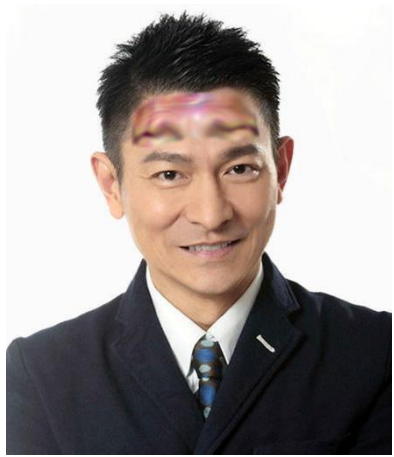

Figure 14 ADVHAT attack sample. Train Andy Lau with an exclusive sticker on his forehead with the texture of another star.

**Mask template experiments**

We mainly test the mask template method: (a) Whether the privacy protection model can be effectively trained through the two networks proposed in this paper to invalidate the BCE and other recognition algorithms; (b) Whether multiple templates can be superimposed to invalidate several recognition algorithms at the same time; (c) Whether the protected image can be restored completely.

*Blinding BCE experiment.* The target algorithm has black box characteristics, so we can only use the same recognition algorithm to train the model at both ends of the network, and then use the same face recognition algorithm to verify the model reasoning. We still use the 207,000 pictures as training and interference samples from CelebA and Andy Lau dataset, and both input and output images are converted to 256×256 pixels.

Principal component analysis (PCA) method is used to reduce dimensionality to accept a two-dimensional space of visual features. From the two dimensional distribution image after the input and output of Mask Template Network and Perturbation Generation Network are continuously printed, the processed image has been different from the original image, as shown in Figure 15. It shows that the method is effective in mathematics.

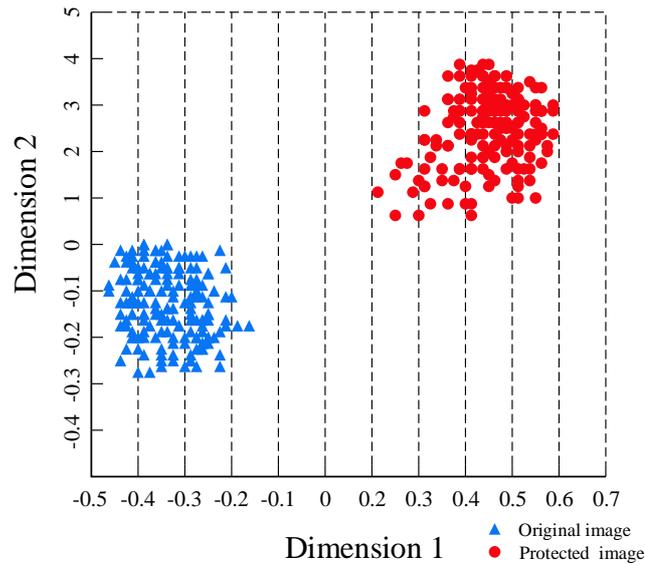

Figure 15 Visual 2D PCA feature space. The Euclidean distance before and after the image passes through the mask template network is obviously different, indicating that it has reached the expectation in principle.

The model generated by 10,000 rounds of training for BCE is named $M_B$. The generated template is directly XOR with the original image. We found that the BCE algorithm cannot recognize the image without additional noise interference (the template will bring some horizontal and vertical texture noise when it is generated). See Figure 16. It shows that this model can effectively protect the key features of the image and meet the requirements of privacy protection.

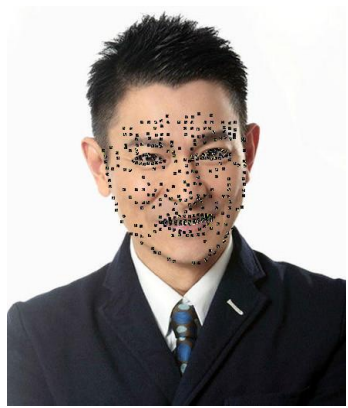

Figure 16 Face mask template on photo. The generated visualization template acts directly on the image, which can effectively reduce the accuracy of commercial algorithms.

Of course, the main content of the image is also difficult to be recognized by the human eyes. To make the content of the template area close to the acceptable range of human eyes, we also need to map some noise to the template, and constantly adjust the noise distribution and threshold to make an image that looks indistinguishable from the original image.

*Human eyes adaptation experiment*. Put the template and noise into the Perturbation Generation Network for continuous adjustment $\sigma$ until a satisfactory image is output. Even if an ideal model is generated in this network, the value of $\sigma$ may need to be adjusted again when implemented. Determining the value of $\sigma$ is a time-consuming task, we can find the best threshold in two stages: The first 7500 rounds of training can make $\sigma = 1$, and the later training is divided into groups every 500 rounds. Referring to the chimp optimization algorithm (ChOA) [42], each feature point is adjusted separately (hunting separately in imitation of the orang-utan population), so that the fastest captured optimal solution can be used as the benchmark interception value, and then sorting and fine-tuning can be carried out after narrowing the range.

If the training has produced serious distortion, the grouping needs to be adjusted to allow more training to participate in ChOA, until $F_c$ and $\sigma$ meet expectations. We find that changing the input of Perturbation Generation Network from Gaussian noise to a specific noise dataset will effectively improve the performance of the network. This dataset can be the pixel values near the feature points corresponding to the mask template of the picture less than 1KB after the scale transformation of CelebA (in fact, these values also conform to the Gaussian distribution). These noises corresponding to the target template area will also produce a smoother effect, see Figure 17. If we can classify the skin colour of the faces in the sample, the effect will be better.

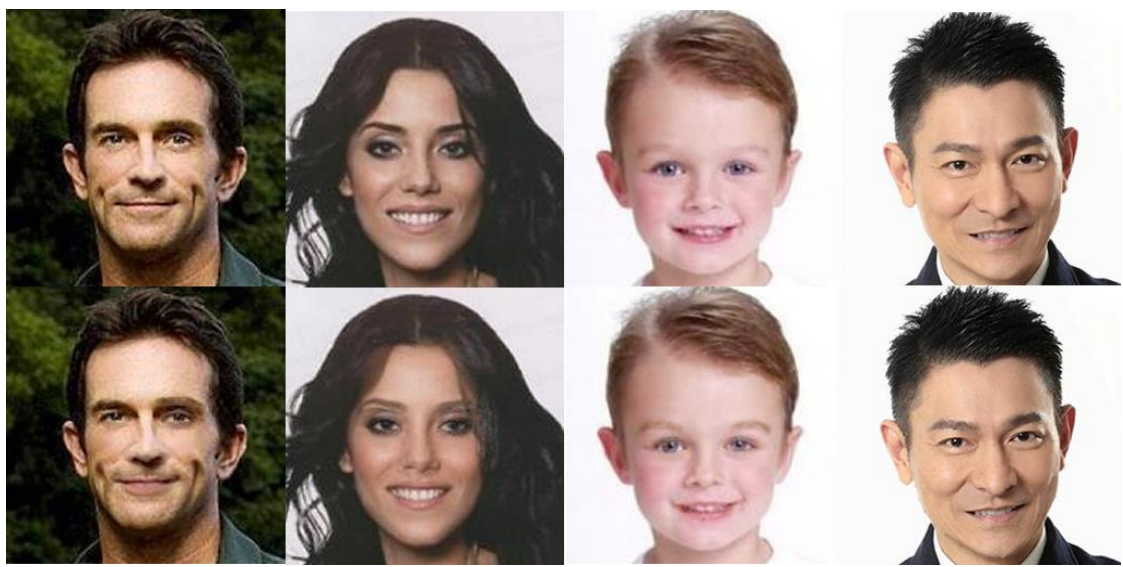

Figure 17 Pairs of original and protected images. The picture above is the original image and below is the protected image, using a specific set of noise looked more natural

*Superposition experiment*. Use the same method of $M_B$ production to generate the models of Azure face, Face++, ArcFace and Dface algorithms in turn, which are recorded as $M_{Az}$, $M_F$, $M_{Ar}$, $M_D$. The test results are shown in Table 2. Experiments are conducted against the target face recognition algorithm, and the output image can deceive them perfectly. The experiments show that this method is suitable for independent face recognition algorithms, and there is a threshold that can blind the target algorithm to protect the image and be recognized by human eyes.

Table 2: Comparison of recognition algorithms training results.

| Modul name | $T_o$ | $F_c$ | $T_r$ | $r_s$ | $|\sigma|$ |
|---|---|---|---|---|---|
| $M_B$ | 0.999 | 0.918 | 0.991 | 0.849 | 0.414 |
| $M_{Az}$ | 0.997 | 0.989 | 0.989 | 0.981 | 0.018 |
| $M_F$ | 0.998 | 0.996 | 0.998 | 0.994 | 0.560 |
| $M_{Ar}$ | 0.986 | 0.936 | 0.985 | 0.891 | 1.121 |
| $M_D$ | 0.965 | 0.999 | 0.957 | 1.034 | 0.002 |

With the help of the feature-representation-transfer[31] mentioned above, we perform algorithm superposition, and there is a catastrophic forgetting (feature confusion) phenomenon of outputting a dithering effect similar to mosaic in the superposition process. Intuitively, it should be substantial differences in the landmarks specifications of these enumerated face recognition algorithms, which may be related to our manual transformation of gradient extremum. The coordinates of the feature points between algorithms will not have 100% conflict. Trying to delete the conflicting feature points will alleviate the impact on the numerical distribution. But if too many coordinates are deleted, the output image may still not work. As the number of compatible recognition algorithms increases, interception parameters need to be adjusted more frequently before outputting the image. The solution is to use τ to specify the target recognition algorithm and process the image for fewer algorithms (less than 3 are recommended) or algorithms with similar principles. The human visual system experience is measured by four indicators after the algorithm is compatible: perfect, normal, unsatisfactory and poor. Because it has subjective color, it cannot be quantified at present. See Figure. 18

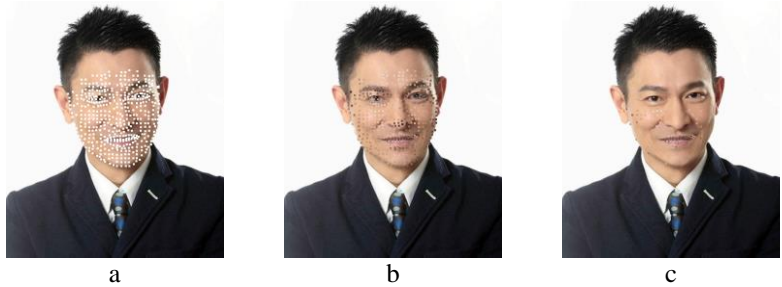

Figure 18 Template transfer experiment. (a) Facial mask template on photo. (b) Feature confusion. (c) Feature point trimming

*Image restoration experiment*. After the network proposed in this paper, the output image results from the XOR operation between the noise and the original image through the mask template, see Figure 19. Therefore, in order to restore the image, we only need to make the output image XOR with the mask template again. The mask template can be saved by image encryption. When the image needs to be decoded, first restore the graphical mask template by decryption, to ensure the security of the mask template key.

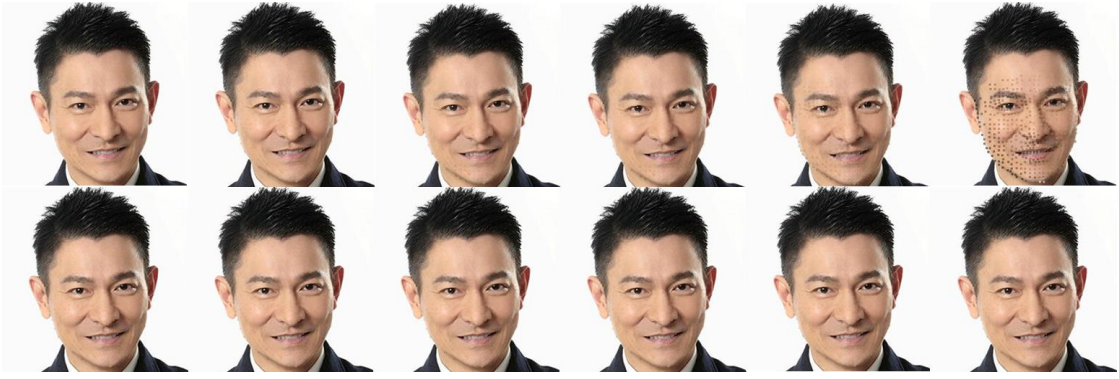

Figure 19 Multi model superposition reduction experiment. From left to right are the images output by $M_{F+Ar}$, $M_{B+Aze}$, $M_{B+Az+F}$, $M_{B+Az+F+Ar}$, $M_{B+Az+F+D}$, $M_{B+Az+F+Ar+D}$ respectively. The image below is restored image.

Input the corresponding models of each group with the protected images to obtain the average error recognition rate $\bar{F}_c$. After restoring the image, calculate the average correct recognition rate $\bar{T}_r$. As shown in Table 3. It records the subjective feelings of human eyes.

Table 3: Comparison of recognition algorithms training results.

| Algorithm name | Perception | $\bar{F}_c$ | $\bar{T}_r$ |
| --- | --- | --- | --- |
| $M_{B+Az+F+Ar+D}$ | Poor | 0.512 | 0.989 |
| $M_{B+Az+F+D}$ | Poor | 0.838 | 0.989 |
| $M_{B+Az+F+Ar}$ | Unsatisfactory | 0.892 | 0.995 |
| $M_{B+Az+F}$ | Normal | 0.977 | 0.998 |
| $M_{B+Aze}$ | Normal | 0.984 | 0.998 |
| $M_{F+Ar}$ | Perfect | 0.996 | 0.997 |

Experiments show that the scheme proposed in this paper can effectively protect privacy and restore in images. It is better than other popular methods.

## Conclusions

In the article, we propose two networks. One is the Mask Template Network, which can generate corresponding mask templates in different recognitions. This template can not only assist the generators to converge quickly, but also restore the protected images as a decoding tool. Another is Perturbation Generation Network whose discriminator uses the target recognition network, its different from GAN. The generator can input Gaussian noise, however, we recommend entering a set of pixels consistent with the corresponding feature point type of the image to be protected. This can significantly reduce the training cost. Compared with the previous methods, ours has certain progress and compatibility. It can make many recognition algorithms invalid without affecting the visual effect of human eyes. The protected images can be restored by mask template. If the template file is encrypted to prevent the key from being leaked and cracked, the data security has been greatly improved.

At present, we have not improved the visual perception of over four recognition algorithms. When adding recognition algorithms with enormous difference in feature space, the image output will become very unstable. The network can only be adjusted by the known recognition algorithm coefficient τ, or some features with large spatial distance can be deleted. The mask template and generator only support fixed size image input and output. We will continue to improve it in the future.

## Data Availability

## Conflicts of Interest

## Funding Statement

This research was partially funded by the Strategic Priority Research Program of Chinese Academy of Sciences (XDB 41020104), and Scientific Research Project of Beijing Educational Committee (KM202110005024).